\documentclass[a4paper,twoside]{article}

\usepackage{epsfig}
\usepackage{subfigure}
\usepackage{calc}
\usepackage{amssymb}
\usepackage{amstext}
\usepackage{amsmath}
\usepackage{amsthm}
\usepackage{comment}
\usepackage{bm}
\usepackage{multirow}
\usepackage{cite}
\usepackage{multicol}
\usepackage{pslatex}
\DeclareMathOperator*{\argmax}{argmax}
\usepackage{apalike}
\usepackage{SCITEPRESS}     

\begin{document}

\title{Efficient and Effective Single-Document Summarizations and A Word-Embedding Measurement of Quality}

\author{\authorname{Liqun Shao, Hao Zhang, Ming Jia and Jie Wang}
\affiliation{Department of Computer Science, University of Massachusetts, Lowell, MA, USA}
\email{\{Liqun\_Shao, Hao\_Zhang, Ming\_Jia\}@student.uml.edu, Jie\_Wang@uml.edu}
}


\abstract{
Our task is to generate an effective summary for a given document with specific realtime requirements. 
We use the softplus function 
to enhance keyword rankings to favor important sentences,
based on which we present a number of summarization algorithms using various
keyword extraction and topic clustering methods. We show 
that our algorithms meet the realtime requirements and yield the best ROUGE recall scores on DUC-02 over all previously-known
algorithms. 
%
%
%
To evaluate the quality of summaries 
without human-generated benchmarks, 
we define a measure called WESM based on word-embedding using 
Word Mover's Distance.
We show that
the orderings of the ROUGE and WESM scores of our algorithms are highly comparable, 
suggesting that WESM may serve as a viable alternative for measuring the quality of a summary.
}

\onecolumn \maketitle \normalsize \vfill

\section{\uppercase{Introduction}}
\label{sec:introduction}
\noindent Text summarization algorithms have been studied intensively and extensively.
An effective summary must
be human-readable and 
convey the central meanings of the original document within
a given length boundary. 
The common approach of unsupervised summarization algorithms
extracts sentences based on importance rankings (e.g., see \cite{DUC02,Mihalcea04,Rose:10,lin:11,ParveenR015}),
where a keyword may also
be a phrase. 
A sentence with a larger number of keywords of higher ranking scores is considered more important 
for extraction.
Supervised algorithms include
CNN and RNN models 
for generating extractive and abstractive summaries (e.g., see \cite{RushCW15,NallapatiZSGX16,ChengL16a}). 
%
%

We were asked to construct a general-purpose text-automation tool to produce, among other things, an effective
summary for a given document with the following realtime requirements: Generate a summary instantly for a document of
up to 2,000 words, under 1 second for
a document of slightly over 5,000 words, and under 3 seconds for a very long document of around 10,000 words.
Moreover, we need to deal with documents of arbitrary topics without knowing what the topics are in advance.
After investigating all existing summarization algorithms,
we conclude that unsupervised single-document summarization algorithms would be the best approach to meeting
our requirements. 

We use topic clusterings to obtain a good topic coverage in the summary when extracting key sentences.
In particular, we first determine which topic a sentence belongs to,
and then 
extract key sentences to cover as many topics as possible within the given length boundary.

Human judgement is the best evaluation of the quality of a summarization algorithm. 
It is a standard practice to run an algorithm over DUC data and
compute the ROUGE recall scores with a set of DUC benchmarks,
which are human-generated summaries for articles of a moderate size.
DUC-02 \cite{DUC02}, in particular, is a small set of benchmarks for single-document summarizations.
When dealing with a large number of documents of unknown topics and various sizes, human judgement may be
impractical, and so
we would like to have an alternative mechanism of measurement 
without human involvement.
Ideally, this mechanism should preserve the same ordering as 
ROUGE over DUC data; namely,
if $S_1$ and $S_2$ are two summaries of the same DUC document 
produced by two algorithms,
and the ROUGE score of $S_1$ is higher than that of $S_2$, then it should also be the case under the new measure.

Louis and Nenkova \cite{Louis:2009} devised an unsupervised method to
evaluate summarization without human models using
common similarity measures: Kullback-Leibler divergence, Jensen-Shannon divergence, and  cosine similarity.
%
%
These measures, 
as well as the information-theoretic similarity measure \cite{Aslam03},
are meant to measure lexical similarities, which are unsuitable for measuring semantic similarities.

Word embeddings such as Word2Vec can
be used to fill this void 
and
we devise WESM (Word-Embedding Similarity Measure)
based on Word Mover's Distance (WMD) \cite{wmd}
to measure word-embedding similarity of the summary and the original document. WESM is meant to evaluate summaries for new datasets when no human-generated benchmarks are available. WESM has an advantage that it can measure the semantic similarity of documents. We show that WESM correlates well with ROUGE on DUC-02. Thus, WESM may be used as an alternative summarization evaluation method when benchmarks are unavailable.



The major contributions of this paper are summarized below:

\vspace*{-3pt}
\begin{enumerate}
\item We present a number of summarization algorithms using topic clustering methods and enhanced keyword rankings by the softplus function,
and show that they meet the realtime requirements and outperform all the previously-known summarization algorithms under
the ROUGE measures over DUC-02.
%
%


\vspace*{-3pt}
\item We propose a new mechanism WESM as
%
an alternative measurement of summary quality when human-generated benchmarks are unavailable.
%
%
\end{enumerate}

The rest of the paper is organized as follows:
We survey in Section \ref{sect:work} unsupervised single-document summarization algorithms.
We present in Section \ref{sect:smethod} the details of our summarization algorithms and describe WESM in Section \ref{sect:emethod}. 
We report the results of extensive experiments in Section \ref{sec:experiments} and conclude the paper in
Section \ref{sect:conclusion}.

\section{\uppercase{Early Work}}
\label{sect:work}

\noindent Early work on single-topic summarizations can be described in the following three categories: keyword extractions, coverage and diversity optimizations,
and topic clusterings.

\subsection{Keyword extractions}

To identify keywords in a document over a corpus of documents, the measure of term-frequency-inverse-document-frequency (TF-IDF) \cite{salton87} is often used.
When document corpora are unavailable,
the measure of word co-occurrences (WCO) can produce a 
comparable performance to TF-IDF over a large corpus of documents \cite{Matsuo:03}.
The methods of TextRank \cite{Mihalcea04} and RAKE (Rapid Automatic Keyword Extraction) \cite{Rose:10}
further refine the WCO method from different perspectives, 
which are also sufficiently fast to become candidates 
for meeting the realtime requirements. 

TextRank 
computes the rank of a word in an undirected, weighted word-graph using a slightly modified PageRank algorithm \cite{Brin:98}.
To construct a word-graph for a given document, first remove stop words and represent each remaining word
as a node, then link two words if they both appear in a sliding window of a small size. Finally,
assign the number of co-occurrences of the endpoints of an edge as a weight to the edge.

RAKE 
first removes stop words using a stoplist, and then generates words (including phrases) using a set of word delimiters and 
a set of phrase delimiters.
For each remaining word $w$, the degree of $w$ is the frequency of $w$ plus the number of co-occurrences of consecutive word pairs $ww'$ and $w''w$ in the document, where $w'$ and $w''$ are
remaining words.
The score of $w$ is the degree of $w$ divided by the frequency of $w$. We note that
the quality of RAKE also depends on a properly-chosen stoplist, which is language dependent.
%

\subsection{Coverage and diversity optimization}

The general framework of selecting sentences gives rise to optimization problems with objective functions
being monotone submodular \cite{lin:11} to promote coverage and diversity.
Among them is an objective function in the form of $L(S) + \lambda R(S)$ with
%
a summary $S$ and a coefficient $\lambda \geq 0$, where
$L(S)$ measures the coverage of the summary and $R(S)$ rewards diversity.
We use SubmodularF to denote the algorithm computing this objective function.
SubmodularF uses TF-IDF values of words in sentences to compute the cosine similarity of two sentences.
While
it is NP-hard to maximize a submodular objective function subject to a summary length constraint, the submodularity allows
a greedy approximation with a proven approximation ratio of $1-1/\sqrt{e}$.
%
%

SubmodularF needs labeled data to train the parameters in the objective function to achieve a better summary and it is intended to work on multiple-document summarizations.
While it is possible to work on a single document without a corpus, we note that the greedy algorithm has at least a quadratic-time complexity and it produces a summary with
low ROUGE scores over DUC-02 (see Section \ref{sec:other}),
and so it would not be a good candidate to meet our needs. This also applies to a generalized objective function
%
%
%
consisting of a submodular component and a non-submodular component \cite{DasguptaKR13}.

\subsection{Topic clusterings}

Two unsupervised approaches to topic clusterings for a given document have been investigated.
One is TextTiling \cite{hearst:97} and the other is LDA (Latent Dirichlet Allocation) \cite{blei:03}.
TextTiling 
represents a topic as a set of consecutive paragraphs in the document.
It merges adjacent paragraphs that belong to the same topic. TextTiling identifies major topic-shifts based on patterns of lexical co-occurrences and distributions.
LDA computes for each word a distribution under a pre-determined number of topics. 
LDA is a computation-heavy algorithm
that incurs a runtime too high to meet our realtime requirements.
TextTiling has a time complexity of almost linear, which meets the requirements of efficiency. 

\subsection{Other algorithms}
\label{sec:other}

Following the general framework of selecting sentences to meet the requirements of topic coverage and diversity,
a number of unsupervised single-document summarization algorithms have been devised.
The most notable is $CP_3$ \cite{parveen2016}, which produces the best ROUGE-1 (R-1), ROUGE-2 (R-2), and ROUGE-SU4 (R-SU4) scores on DUC-02
among all early algorithms,
including
Lead \cite{ParveenR015}, DUC-02 Best, TextRank, LREG \cite{ChengL16a},
Mead \cite{radev2004}, $ILP_{phrase}$ \cite{woodsend2010}, URANK \cite{wan2010}, UniformLink \cite{WanX10}, Egraph + Coherence \cite{Parveen015},
Tgraph + Coherence (Topical Coherence for Graph-based Extractive Summarization) \cite{ParveenR015},
NN-SE \cite{ChengL16a}, and SubmodularF.

$CP_3$ maximizes importance, non-redundancy, and pattern-based coherence of sentences to generate a coherent summary using ILP. 
It computes 
the ranks of selected sentences for the summary by the Hubs and Authorities algorithm (HITS) \cite{kleinberg1999}, 
and ensures that each selected sentence has unique information. 
It then uses mined patterns to extract sentences if the connectivity among nodes in the projection graph matches the connectivity among nodes in a coherence pattern. Because of space limitation, we omit the descriptions of the other algorithms.

Table \ref{CP3} shows the comparison results, where the results for SubmodularF
is obtained using
the best parameters trained on DUC-03 \cite{lin:11}.
Thus, to demonstrate the effectiveness of our algorithms, we will compare our algorithms with only $CP_3$ 
over DUC-02.
\begin{table}[h]
\begin{center}
\caption{\label{CP3} ROUGE scores (\%) on DUC-02 data.}
\begin{tabular}{l|c|c|c}
\hline
\bf Methods & \bf R-1 & \bf R-2 & \bf R-SU4 \\
\hline
Lead & 45.9 & 18.0 & 20.1 \\
DUC 2002 Best & 48.0 & 22.8 & \\
TextRank & 47.0 & 19.5 & 21.7 \\
LREG & 43.8 & 20.7 & \\
Mead & 44.5 & 20.0 & 21.0 \\
$ILP_{phrase}$ & 45.4 & 21.3 & \\
URANK & 48.5 & 21.5 & \\
UniformLink & 47.1 & 20.1 &   \\
Egraph + Coh. & 48.5 & 23.0 & 25.3  \\
Tgraph + Coh. & 48.1 & 24.3 & 24.2 \\
NN-SE & 47.4 & 23.0 & \\
SubmodularF & 39.6 & 16.9 & 17.8 \\
$\bm{CP_3}$& \bf 49.0 & \bf 24.7 & \bf 25.8  \\
\hline
\end{tabular}

\end{center}
\end{table}

Solving ILP, however, is
time consuming even on documents of a moderate size, for ILP is NP-hard.
%
Thus, $CP_3$ does not meet the requirements of time efficiency. We will need to investigate new methods.

\section{\uppercase{Our Methods}}

\label{sect:smethod}

\noindent We use TextRank and RAKE to obtain initial ranking scores of keywords, and
use the softplus function \cite{softplus}
\begin{equation}\label{eq1}
sp(x) = \ln (1+e^x)
\end{equation}
to
enhance keyword rankings to favor sentences that are more important. 

\subsection{Softplus ranking}

Assume that after filtering, a sentence $s$ consists of $k$ keywords $w_1, \cdots, w_k$, and $w_i$ has a ranking score $r_i$
produced by TextRank or RAKE. 
Following Shao and Wang ~\cite{Shao:16}, we use their central sentence extraction algorithm for ranking sentences by importance as $\mbox{Rank}(s)$.
We can rank $s$ using one of the following two methods:
\begin{equation}\label{eq2}
{Rank}(s) = \sum_{i=1}^{k} r_i
\end{equation}
\begin{equation}\label{eq3}
{Rank}_{sp}(s) = \sum_{i=1}^{k} sp(r_i)
\end{equation}
Let DTRank (Direct TextRank) and ETRank (Enhanced TextRank) denote
the methods of ranking sentences using, respectively,
$\mbox{Rank}(s)$ and $\mbox{Rank}_{sp}(s)$ over TextRank keyword rankings,
and
%
DRAKE (Direct RAKE) and ERAKE (Enhanced RAKE)
to denote the methods of ranking sentences
using, respectively, $\mbox{Rank}(s)$ and $\mbox{Rank}_{sp}(s)$ over RAKE keyword rankings.

%
%

\begin{figure}[h]
\centering
\includegraphics[width=2.5in]{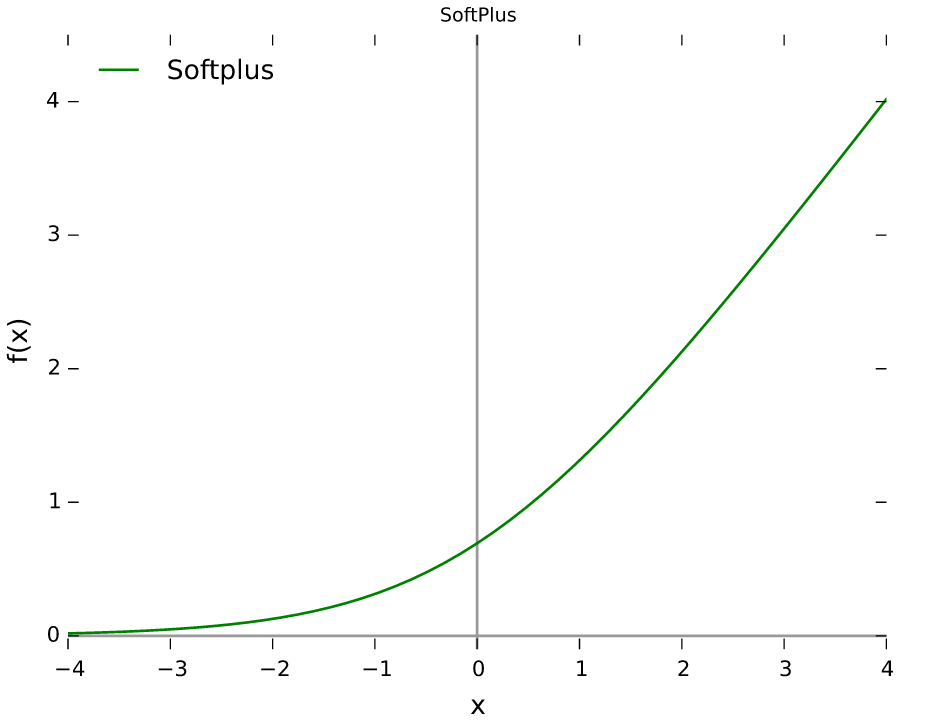}
\caption{Softplus function $\ln(1+e^x)$.} 
\label{fig:softplus}
\end{figure}

The softplus function is helpful because 
when $x$ is a small positive number, $sp(x)$ increases the value of $x$ significantly (see Figure \ref{fig:softplus})
and when $x$ is large, $sp(x) \approx x$.
In particular,
given two sentences $s_1$ and $s_2$, suppose that $s_1$ has a few keywords with high rankings and the rest of the keywords with low rankings,
while $s_2$ has medium rankings for almost all the keywords. In this case, we would consider $s_1$ more important than $s_2$. However,
we may end up with $\mbox{Rank}(s_1) < \mbox{Rank}(s_2)$.
To illustrate this using a numerical example, assume that $s_1$ and $s_2$ each consists of 5 keywords, with
original scores (sc) and softplus scores (sp) given in the following table \ref{example}.

\medskip


\noindent
\begin{table}[h]
\begin{center}
\caption{\label{example} Numerical examples with given sc and sp scores.}
\begin{tabular}{l||c|c|c|c|c||c}
\hline
$s_1$ & $w_{11}$ & $w_{12}$ & $w_{13}$ & $w_{14}$ & $w_{15}$&   Rank \\
\hline
sc & 2.6 & 2.2 & 2.1 & 0.3 & 0.2 & 7.4\\
sp & 2.67 & 2.31 & 2.22 & 0.85& 0.80 & \bf 8.84\\
\hline
$s_2$ & $w_{21}$ & $w_{22}$ & $w_{23}$ & $w_{24}$ & $w_{25}$ & \\
\hline
sc & 1.6 & 1.5 & 1.5 & 1.5 & 1.4  & \bf 7.5  \\
sp & 1.78& 1.70& 1.70& 1.70& 1.62&  8.51 \\
\hline
\end{tabular}
\end{center}
\end{table}

\medskip
Sentence $s_1$ is more important than $s_2$
because it contains three keywords of much higher ranking scores than those of $s_2$.
However, $s_2$ will be selected without using softplus. After using softplus, $s_1$ is selected as it should be.

For a real-life example, consider the following two sentences from an article in DUC-02:
\vspace*{-1pt}
\begin{itemize}
\item[$s_1$:] {\small\sf Hurricane Gilbert swept toward Jamaica yesterday with 100-mile-an-hour winds, and officials issued warnings to residents on the southern coasts of the Dominican Republic, Haiti and Cuba.}
\vspace*{-1pt}\item[$s_2$:] {\small\sf Forecasters said the hurricane was gaining strength as it passed over the ocean and would dump heavy rain on the Dominican Republic and Haiti as it moved south of Hispaniola, the Caribbean island they share, and headed west.}
\end{itemize}
We consider $s_1$ more important as it
specifies the name, strength, and direction of the hurricane, the places affected, and the official warnings.
Using TextRank to compute
keyword scores, we have $\mbox{Rank}(s_1) = 1.538 < \mbox{Rank}(s_2) = 1.603$, which returns a less important sentence $s_2$. After computing
softplus,
we have $\mbox{Rank}_{sp}(s_1) = 8.430 > \mbox{Rank}_{sp}(s_2) = 7.773$; the more important sentence $s_1$ is selected.

Note that not any exponential function would do the trick. What we want is a function to return roughly the same value as the input when the input is large, and a significantly larger value than the input when the input is much less than 1. The softplus function meets this requirement.

\subsection{Topic clustering schemes}

We consider four topic clustering schemes: TCS, TCP, TCTT, and TCLDA.
\begin{enumerate}
\item TCS 
selects sentences without checking topics.
\item TCP 
treats each paragraph as a separate topic.
\item TCTT 
partitions a document into a set of multi-paragraph 
segments using TextTiling.
\item TCLDA 
computes a topic distribution for each word using LDA. 
We set the number of topics from 5 to 8 depending on the length of the document.
Assume that a document contains $K$ topics ($5 \leq K \leq 8$) and the topic $j$ consists of $k_j$ words $w_{1j}, \cdots, w_{k_j,j}$, where $1 \leq j \leq K$
and
$w_{ij}$ has a probability $p_{ij} > 0$. For a document with $n$ sentences $s_1, \cdots, s_n$,
we use the following maximization to determine which topic $t_z$ the sentence $s_z$ belongs to ($1 \leq t \leq K$):
\begin{equation} \label{eq4}
t_z = \argmax_{1 \leq j\leq k}\bigg(\prod_{i:w_{ij} \in s_z} {p_{ij}}\bigg)
\end{equation}
\end{enumerate}

\subsection{Summarization algorithms}
\label{ssect:sa}

The length of a summary may be specified by users,
either as a number of words or as a percentage of the number of characters of the original document.
By a ``30\% summary'' we mean that the number of characters of the summary does not exceed 30\% of that of the original document.

Let $L$ be the summary length (the total number of characters) specified by the user and $S$ a summary.
If $S$ consist of $m$ sentences
$s_1, \cdots, s_m$, and the number of characters of $s_i$ is $\ell_i$, then the following inequality must hold:
$\sum_{i=1}^m \ell_i \leq L.$

Depending on which sentence-ranking algorithm and which topic-clustering scheme to use, we have eight combinations
using ETRank and ERAKE, and eight combinations using DTRank and DRAKE, shown in Table \ref{algorithm_names}.
%
For example, ET3Rank (Enhanced TextTiling TRank) means to use $\mbox{Rank}_{sp}(s)$ to rank sentences and
TextTiling to compute topic clusterings, and
%
T2RAKE (TextTiling RAKE)
means to use $\mbox{Rank}(s)$ rank sentences over RAKE keywords and TextTiling
to compute topic clusterings. 

\begin{table}[h]
\begin{center}
\caption{\label{algorithm_names} Description of all the Algorithms with different sentence-ranking (S-R) and topic-clustering (T-C) schemes.}
\begin{tabular}{l|c|c}
\hline
\bf Methods & \bf S-R & \bf T-C \\
\hline
ESTRank & ETRank & TCS \\
EPTRank & ETRank & TCP \\
ET3Rank & ETRank & TCTT \\
ELDATRank & ETRank & TCLDA \\
\hline
ESRAKE & ERAKE & TCS \\
EPRAKE & ERAKE & TCP \\
ET2RAKE & ERAKE & TCTT \\
ELDARAKE & ERAKE & TCLDA \\
\hline
STRank & DTRank & TCS \\
PTRank & DTRank & TCP \\
T3Rank & DTRank & TCTT \\
LDATRank & DTRank & TCLDA \\
\hline
SRAKE & DRAKE & TCS \\
PRAKE & DRAKE & TCP \\
T2RAKE & DRAKE & TCTT \\
LDARAKE & DRAKE & TCLDA \\
\hline
\end{tabular}
\end{center}
\end{table}

All algorithms follow the following procedure for selecting sentences:

\vspace*{-2pt}
\begin{enumerate}
\item Preprocessing phase
\begin{enumerate}
\vspace*{-1pt}\item Identify keywords and compute the ranking of each keyword.
\vspace*{-1pt}\item Compute the ranking of each sentence.
\end{enumerate}
\vspace*{-2mm}
\item Sentence selection phase
\begin{enumerate}
\vspace*{-1pt}\item Sort the sentences in descending order of their ranking scores. 

\vspace*{-1pt}\item Select sentences one at a time with a higher score to a lower score.
Check if the selected sentence $s$ belongs to the known-topic set (KTS) according to the underlying
topic clustering scheme, where KTS is a set of topics from sentences placed in the summary so far. If $s$ is in KTS, then discard it; otherwise, place $s$ into the summary and its topic into KTS.
\vspace*{-1pt}\item Continue this procedure until the summary reaches its length constraint.

\vspace*{-1pt}\item If the number of topics contained in the KTS is equal to the number of topics in the document,
empty KTS and repeat the procedure from Step 1. 
\end{enumerate}
\end{enumerate}

\begin{figure*}[t]
\includegraphics[width=6in]{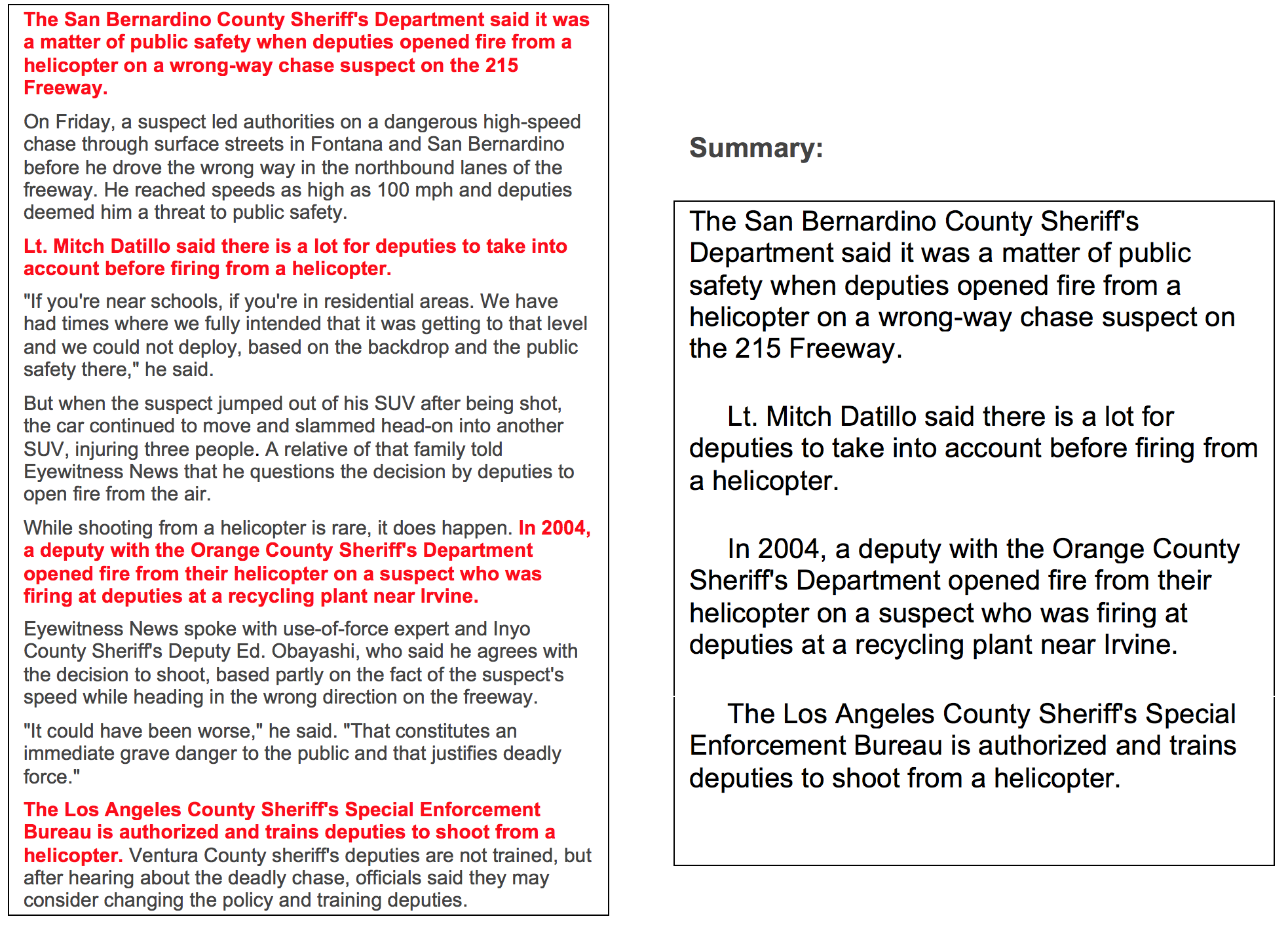}
\caption{An example of 30\% summary of an article in NewsIR-16 by ET3Rank, where the
original document is on the left and the summary is on the right.}
\label{fig:fig5}
\end{figure*}

Figure \ref{fig:fig5} shows an example of 30\% summary generated by ET3Rank on an article in NewsIR-16.

\section{\uppercase{A Word-Embedding Measurement of Quality}}
\label{sect:emethod}

\noindent Word2vec \cite{mikolov13,mikolov2013}
is an NN model that learns a vector representation for each word contained in a corpus of documents.
The model consists of an input layer, a projection layer, and an output layer to predict nearby words in the context. In particular,
a sequence of $T$ words $w_1, \cdots, w_T$ are used to train a Word2Vec model for maximizing the probability of neighboring words:
\begin{equation} \label{eq5}
{\frac{1}{T}\sum_{t=1}^{T}{\sum_{j\in b(t)}{\log p(w_j|w_t)}}}
\end{equation}
where $b(t) = [t-c, t + c]$ is the set of center word $w_t$'s neighboring words, $c$ is the size of the training context, and $p(w_j|w_t)$ is defined by the softmax function.
Word2Vec can learn complex word relationships if it trains on a very large data set.

\subsection{Word Mover's Distance}
\label{ssec:wmd}
Word Mover's Distance (WMD) \cite{wmd} uses Word2Vec as a word embedding representation method.
It measures the dissimilarity between two documents and calculates the minimum cumulative distance to ``travel'' from the embedded words of one document to the other. Although two documents may not share any words in common, WMD can still measure the semantical similarity by considering their word embeddings, while other bag-of-words or TF-IDF methods only measure the similarity by the appearance of words. A smaller value of WMD indicates that the two sentences are more similar.

\subsection{A word-embedding similarity measure}

Based on WMD's ability of measuring the semantic similarity of documents, 
we propose a summarization evaluation measure WESM (Word-Embedding Similarity Measure). 
Given two documents $D_1$ and $D_2$, let $\mbox{WMD}(D_1,D_2)$ denote
the distance of $D_1$ and $D_2$.
Given a document $D$, assume that it consists of $\ell$ paragraphs $P_1, \cdots, P_\ell$.
Let $S$ be a summary of $D$.
We compare the word-embedding similarity of a summary $S$ with $D$ using WESM$(S,D)$ as follows:
\begin{equation} \label{eq6}
\mbox{WESM}(S,D) = \frac{1}{\ell}\sum_{i=1}^{\ell} \frac{1}{1 + \mbox{WMD}(S,P_i)}
\end{equation}
The value of WESM$(S,D)$ 
is between 0 and 1. Under this measure, the higher the WESM$(S,D)$ value, the more similar $S$ is to $D$. 

\section{\uppercase{Numerical Analysis}}
\label{sec:experiments}

%

\noindent We evaluate the qualities of summarizations using the DUC-02 dataset \cite{DUC02}
and the NewsIR-16 dataset \cite{Signal1M2016}.
DUC-02 consists of 60 reference sets, each of which consists of a number of documents, single-document summary benchmarks, and multi-document abstracts/extracts.
The common ROUGE recall measures of ROUGE-1, ROUGE-2, and ROUGE-SU4 are used to compare the quality of summarization algorithms over DUC data.
NewsIR-16 consists of 1 million articles 
from English news media sites and blogs.

We use various software packages to implement TextRank (with window size = 2) \cite{TRurl}, RAKE \cite{RAurl}, TexTiling \cite{TTurl}, LDA and Word2Vec \cite{gen}.
%
%

We use the existing Word2Vec model trained on English Wikipedia \cite{wiki},
which consists of 3.75 million articles formatted in XML. The reason to choose this dataset is
for its large size and the diverse topics it covers.

%

\subsection{ROUGE evaluations over DUC-02}

As mentioned before, we use 
$CP_3$ to cover all previously known algorithms for the purpose of comparing qualities of summaries, as
$CP_3$ produces the best results among them. 

Among all the algorithms we devise, we only present those with at least one ROUGE recall score better than or equal to the corresponding score of $CP_3$, 
identified in bold 
(see Table \ref{duc}).
Also shown in the table is the average of the three ROUGE scores (R-AVG). We can see that
ET3Rank is the winner, followed by T2RAKE; both are superior to $CP_3$.
Moreover, ET2RAKE offers the highest
ROUGE-1 score of 49.3.
%
%
\begin{table}[h]
\begin{center}
\caption{\label{duc} ROUGE scores (\%) on DUC-02 data.}
\begin{tabular}{l|c|c|c||c}
\hline
\bf Methods & \bf R-1 & \bf R-2 & \bf R-SU4 & \bf R-AVG \\
\hline
$CP_3$& 49.0 & 24.7 & 25.8 & 33.17 \\
\hline
\bf ET3Rank & \bf 49.2 & \bf 25.6 & \bf 27.5 & \bf 34.10 \\
ESRAKE & \bf 49.0 & 23.6 & \bf 26.1 &  32.90 \\
ET2RAKE & \bf 49.3 & 21.4 & 24.5  & 31.73\\
PRAKE & \bf 49.0 & 24.5 & 25.3 & 32.93\\
\bf T2RAKE & \bf 49.1 & \bf 25.4 & \bf 25.8 & \bf 33.43 \\
\hline
\end{tabular}
\end{center}
\end{table}

\subsection{WESM evaluations over DUC-02 and NewsIR-16}

%
Table \ref{results1} shows the evaluation results on DUC-02 and NewsIR-16 using WESM based on the Word2Vec model trained on English Wikipedia.
The first number in the third row is the average score on all benchmark summaries in DUC-02.
For the rest of the rows, each number is the average
score of summaries produced by the corresponding algorithm for all documents in DUC-02 and NewsIR-16.
The size constraint of a summary on DUC-02 for each document is the same as that of the corresponding DUC-02 summary benchmark.

For NewsIR-16, we select at random 1,000 documents from NewsIR-16 and remove the title, references, and other unrelated content from each article.
Based on an observation that a 30\% summary allows for a good summary,
we compute 30\% summaries of these articles using each algorithm.
%


\begin{table}[h]
\begin{center}
\caption{\label{results1} Scores (\%) over DUC-02  and NewsIR-16 under WESM trained on English-Wikipedia.}
\begin{tabular}{l|c|c}
\hline
\bf Datasets & \bf DUC-02 & \bf NewsIR-16 \\ 
\hline
Benchmarks & 3.021  & \\
\hline

ET3Rank &\bf 3.382 &\bf 2.002 \\ 
ESRAKE      &3.175 & 1.956  \\ 
ET2RAKE     &3.148 &1.923   \\ 
PRAKE       &3.150 &1.970   \\ 
T2RAKE      &3.247 &1.990   \\ 
\hline
\end{tabular}
\end{center}
\end{table}
It is expected that scores of our algorithms are better than the score for benchmarks under each measure, for the benchmarks often use different words not in the original documents, and hence would
have smaller similarities.

\subsection{Normalized $L_1$-norm}

We would like to determine if WESM is a viable measure. From our experiments, we know that the all-around best algorithm ET3Rank, the second best
algorithm T2RAKE, and ET2RAKE remain the same positions under R-AVG over DUC-02 and under WESM over both DUC-02 and NewsIR-16 (see Table \ref{ordering}),
ESRAKE and PRAKE remain the same positions under R-AVG over DUC-02 and under WESM over NewsIR-16, while ESRAKE and PRAKE only differ by one place under R-AVG and WESM over DUC-02.
\begin{table}[h]
\begin{center}
\caption{\label{ordering} Orderings of R-AVG scores over DUC-02 and
WESM scores over DUC-02 and NewsIR-16.}
\begin{tabular}{l|c|c|c}
\hline
\multirow{2}{*}{\bf Methods} &
\bf R-AVG  & \multicolumn{2}{|c}{\bf WESM} \\
\cline{2-4}
& DUC-02 & DUC-02 & NewsIR-16 \\
\hline
\bf ET3Rank     &\bf 1  &\bf 1 &\bf 1  \\ 
ESRAKE          &4  &3 &4  \\ 
\bf ET2RAKE     &\bf 5  &\bf 5 &\bf 5  \\ 
PRAKE          &3  &4 &3  \\ 
\bf T2RAKE      &\bf 2  &\bf 2 &\bf 2  \\ 
\hline
& ${\bm O}_1$ & ${\bm O}_2$ & ${\bm O}_3$ \\
\hline
\end{tabular}
\end{center}
\end{table}

Next, we compare the ordering of the R-AVG scores and the WESM scores over DUC-02. For this purpose, we use the normalized $L_1$-norm to compare the distance of two orderings. Let ${\bm X} = (x_1, x_2, \cdots, x_k)$ be a sequence of $k$ objects, where
each $x_i$ has two values $a_i$ and $b_i$ such that
$a_1, a_2, \ldots, a_k$ and $b_1,b_2,\ldots, b_k$ are, respectively, permutations of $1,2,\ldots,k$.
Let
\[
D_k= \sum_{i=1}^k |(k-i+1)-i|,
\]
which is the maximum distance two permutations can possibly have. Then
the normalized $L_1$-norm of ${\bm A} = (a_1, a_2, \cdots, b_k)$ and ${\bm B} = (b_1, b_2, \cdots, b_k)$ is defined by
$$||{\bm A}, {\bm B}||_1 = \frac{1}{D_k}\sum_{i=1}^k |a_i - b_i|.$$
Table \ref{ordering} shows the orderings of the R-AVG scores over DUC-02 and WESM scores over DUC-02 and NewsIR-16 (from Tables \ref{duc} and \ref{results1}).

It is straightforward to see that $D_5 = 12$,
$||{\bm O}_1, {\bm O}_2||_1 = ||{\bm O}_2, {\bm O}_3||_1 = 2/12 = 1/6$ and $||{\bf O}_1,{\bf O}_3||_1 = 0$.
This indicates that WESM and ROUGE are highly comparable over DUC-02 and NewsIR-16,
and the orderings of WESM on different datasets, while with larger spread, are
still similar.

\subsection{Runtime analysis}

\begin{figure}[h]
\centering
\includegraphics[width=3.3in]{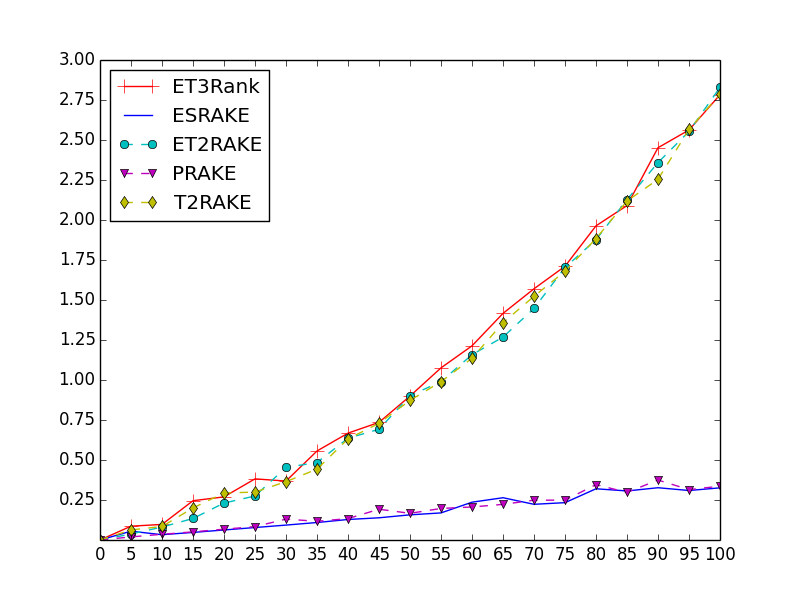}
\caption{Runtime analysis, where the unit on the x-axis is 100 words and the unit of the y-axis is seconds.} \label{fig:runtime}
\end{figure}

We carried out runtime analysis through experiments on a computer with a 3.5 GHz Intel Xeon CPU E5-1620 v3. 
We used a Python implementation of our summarization algorithms. 
Since DUC-02 are short, all but LDA-based algorithms run in about the same time.
To obtain a finer distinction, we ran our experiments on NewsIR-16. Since the average size of NewsIR-16 articles is 405 words,
we selected at random a number of articles from NewsIR-16 and merged them to generate a new article.
For each size from around 500 to around 10,000 words, with increments of 500 words, we selected at random 100 articles and
computed the average runtime of different algorithms to produce 30\% summary (see Figure \ref{fig:runtime}).
We note that the time complexity of each of our algorithms incurs mainly in
the preprocessing phase; 
the size of summaries in the sentence selection phase only incur minor fluctuations of computation time, and
so it suffice to compare the runtime for producing 30\% summaries.

We can see from Figure \ref{fig:runtime} that ESRAKE and PRAKE incur about the same linear time and they are extremely fast.
Also, ET3RANK, ET2RAKE, and T2RAKE incur about the same time. While the time is higher because of the use of TextTiling and
is closed to being linear, it meets the realtime requirements. For example, for a document of up to 3,000 words, over
3,000 but less than 5,500 words, and 10,000 words, respectively,
the runtime of ET3Rank is under 0.5, 1, and 2.75 seconds.
%

The runtime of SubmodularF is acceptable for documents of moderate sizes (not shown in the paper); but for a document of about 10,000 words, the runtime is close to 4 seconds.
LDA-based algorithms is much higher. For example, LDARAKE incurs about 16 seconds for a document of
about 2,000 words, about 41 seconds for a document of about 5,000 words, and about 79 seconds for a document of about 10,000 words.

%

\section{\uppercase{Conclusions}}
\label{sect:conclusion}
\noindent We presented a number of unsupervised single-document summarization algorithms for generating effective summaries in realtime and
a new measure based on word-embedding similarities to evaluate the quality of a summary. We showed that ET3Rank is the best all-around algorithm. A web-based summarization tool using ET3Rank and T2RAKE will be made available to the public.

To further obtain better topic clusterings efficiently, we plan to extend TextTiling over non-consecutive paragraphs.
To obtain a better understanding of word-embedding similarity measures, we plan to compare WESM with human evaluation and
other unsupervised methods including those devised by Louis and Nenkova \cite{Louis:2009}.
We also plan to
explore new ways to measure summary qualities
without human-generated benchmarks.

\section*{\uppercase{Acknowledgements}}

\noindent
We thank the members of the Text Automation Lab at UMass Lowell for their support and fruitful discussions.




\vfill
\bibliographystyle{apalike}
{\small
\bibliography{example}}


\vfill
\end{document}